\documentstyle[12pt]{article}
\begin{document}

\title{$SO(10)$ and Large $\nu_{\mu}-\nu_{\tau}$ Mixing}

\author{{\bf S.M. Barr}\\ Bartol Research Institute, Univ. of Delaware\\
Newark, DE 19716 \\ and \\ {\bf C.H. Albright}\\ Fermi National Accelerator
Laboratory\\ Batavia, IL 60510}

\maketitle

\begin{abstract}
A general approach to understanding the large mixing seen in
atmospheric neutrinos is explained, as well as a highly predictive $SO(10)$
model which implements this approach. It is also seen how bimaximal
mixing naturally arises in this scheme.
\end{abstract}

\vspace{0.5cm}

\begin{center}

{\it Talk given at NNN99, SUNY Stony Brook, Sept. 22-24, 1999}

\end{center}

\vspace{1cm}

The problem of neutrino mixing should not be looked at in
isolation, but as part of the larger ``fermion mass problem",
i.e. the problem of understanding the masses and mixings of the
quarks and leptons. The fermion mass problem is very old; serious efforts 
at model building go back more than twenty-five years, and in that time
hundreds of models of quark and lepton masses have been published.
However, we are now in a new and more hopeful situation, for neutrino
oscillations give us precious new clues in the search for the right
theory as well as new ways of testing theories experimentally.

The most significant new clue, perhaps, is the largeness of the
mixing of $\nu_{\mu}$ (presumably with $\nu_{\tau}$) seen at SuperK.
What is this clue telling us? I will explain one 
idea for what it might be telling us which is
based on certain features of grand unification. I will then 
mention and very briefly discuss a concrete model that incorporates this
idea. (Actually, the model came first, and only then was it noticed
that the model contains an attractive explanation of
the large $\nu_{\mu}-\nu_{\tau}$ mixing!) At the end I will
address the question whether
large $\nu_e$ mixing is compatible with $SO(10)$ in a simple
way. The answer is yes.

Large $\nu_{\mu}-\nu_{\tau}$ mixing came as a surprise. 
The puzzle is that the mixing of the second and third families
is small for the quarks ($V_{cb} \cong 0.04$) and large for the leptons
($U_{\mu  3} \cong 1/\sqrt{2} \cong 0.7$). This is puzzling because both 
grand unification and flavor symmetry, the two most promising ideas for
explaining fermion masses, tend to relate the quark and lepton parameters.
Actually, there are {\it two} puzzles: the mixing of the
second and third families is {\it too small for the quarks} and {\it too 
big for the leptons}, in a sense that I will now explain. 

Many models are 
based on the old idea of
Weinberg, Wilczek and Zee, and Fritzsch.$^1$ Looking at only the first two
families, in the late 1970's they posited simple textures of the form

\begin{equation}
\overline{u_{iR}} U_{ij} u_{jL} =
(\overline{u_R}, \overline{c_R}) \left( \begin{array}{cc}
0 & a \\ a & b \end{array} \right) \left( \begin{array}{c}
u_L \\ c_L \end{array} \right) 
\end{equation}

\noindent
and

\begin{equation}
\overline{d_{iR}} D_{ij} d_{jL} =
(\overline{d_R}, \overline{s_R}) 
\left( \begin{array}{cc} 0 & a' \\ a' & b' \end{array} \right)
\left( \begin{array}{c} d_L \\ s_L \end{array} \right).
\end{equation}

\noindent
The crucial features of these matrices are that they are {\it
hierarchical} (i.e. $a \ll b$ and $a' \ll b'$) and {\it symmetric}.
The eigenvalues of the down quark matrix 
are $m_s \cong b'$, and $m_d \cong - a'^ 2/b'$, and the rotation
angle needed to diagonalize it is $\theta_{ds} \cong a'/b' \cong
\sqrt{m_d/m_s}$, with analogous formulas for the up quarks. 
Thus the Cabibbo angle is given by

\begin{equation}
V_{us} \cong \sqrt{m_d/m_s} - e^{i \phi_{12}} \sqrt{m_u/m_c}. 
\end{equation}

\noindent
Since $V_{us} \cong 0.21$, $\sqrt{m_d/m_s} \cong 0.21$, and $\sqrt{m_u/m_c}
\cong 0.07$, this relation works well for $\phi_{12} \sim \pi/2$.
If one chooses analogous hierarchical and symmetric forms for the
lepton mass matrices one obtains

\begin{equation}
U_{e2} \cong \sqrt{m_e/m_{\mu}} - e^{i \phi'_{12}} \sqrt{m_{\nu_e}/
m_{\nu_{\mu}}}.
\end{equation}

\noindent
This also can work reasonably well if the small angle MSW solution to the
solar neutrino problem is correct, as then $U_{e2} \sim 0.04$, and 
$\sqrt{m_e/m_{\mu}} \cong 0.07$.

If one supposes that the full $3 \times 3$ mass matrices are
symmetrical and hierarchical, one obtains similar predictions for
the mixing of the second and third families, For example, extending
the Weinberg-Wilczek-Zee-Fritzsch pattern to the third family, 
as Fritzsch did,$^2$
one finds

\begin{equation}
V_{cb} \cong \sqrt{m_s/m_b} - e^{i \phi_{23}} \sqrt{m_c/m_t}, 
\end{equation}

\noindent
and

\begin{equation}
U_{\mu 3} \cong \sqrt{m_{\mu}/m_{\tau}} - e^{i \phi'_{23}}
\sqrt{m_{\nu_{\mu}}/m_{\nu_{\tau}}}.
\end{equation}

\noindent
Since $\sqrt{m_s/m_b} \cong 0.14$, and $\sqrt{m_c/m_t} \cong 0.04$, 
one sees that the experimental value of the quark mixing $V_{cb}$ 
($\cong 0.04$) is about a factor 
of 3 smaller than the Fritzschian expectation. On the other hand, since
$\sqrt{m_{\mu}/m_{\tau}} \cong 0.24$, and $\sqrt{m_{\nu_{\mu}}/
m_{\nu_{\tau}}}$ may be presumed to be small, one sees that
the experimental value of the lepton mixing $U_{\mu 3}$ ($\cong 0.7$)
is about a factor of 3 larger than the Fritzschian expectation.

What we have argued in several papers$^{3,4,5}$ is that the trouble 
with such Fritzschian textures when applied to the heavier two families is 
that they are based on symmetric matrices. Let us see what happens if there  
are instead highly asymmetric or, as we have called them, ``lopsided"  
textures. Consider a toy example with matrices

\begin{equation}
\overline{d_{iR}} D_{ij} d_{jL} =
m(\overline{d_R}, \overline{s_R}, \overline{b_R}) 
\left( \begin{array}{ccc}
- & - & - \\ 
- & 0 & \sigma \\ - & \epsilon & 1 \end{array} \right) 
\left( \begin{array}{c}
d_L \\ s_L \\ b_L \end{array} \right), 
\end{equation}

\begin{equation}
\overline{\ell_{iR}} L_{ij} \ell_{jL} =
m(\overline{e_R}, \overline{\mu_R}, \overline{\tau_R})
\left( \begin{array}{ccc} - & - & - \\
- & 0 & \epsilon \\ - & \sigma & 1 \end{array} \right)
\left( \begin{array}{c} e_L \\ \mu_L \\ \tau_L \end{array} \right),
\end{equation}

\noindent
where $\epsilon \ll \sigma \sim 1$. We are interested at the moment 
in the heavier two families, so we do not write the entries of the first
row and column, which are assumed to be small. The important feature of 
these matrices is the large, lopsided off-diagonal entry ($\sigma$).
For the mass matrices of the up quarks and neutrinos we assume that
there is no such large off-diagonal element.
Note another very important feature of these matrices, which is
that the charged lepton mass matrix $L$
is the {\it transpose} of the down quark
mass matrix $D$. $L = D^T$ 
is a ``minimal $SU(5)$" relation. The realistic
model we will discuss later is based on $SO(10)$, nevertheless the $SU(5)$
subgroup of $SO(10)$ in that model relates $L$ to $D^T$.
The reason $L$ is related to $D^T$ is simple: $SU(5)$ unifies the $\ell^-_L$
with the $d_R$ in $\overline{{\bf 5}}$ multiplets
and the $\ell^-_R$ with the $d_L$ in ${\bf 10}$ multiplets. Thus
the mass matrix of the charged leptons is related to that of the
down quarks only up to a left-right transposition. As we shall see,
this feature allows a simple explanation of the double puzzle of
$U_{\mu 3}$ and $V_{cb}$. 

The point is that the observed mixings are of {\it left-handed}
fermions. Specifically, we may write $U_{\mu 3} \cong \theta_{\mu \tau}^{left}
- \theta_{\nu_{\mu} \nu_{\tau}}^{left}$, and $V_{cb} \cong
\theta_{sb}^{left} - \theta_{ct}^{left}$. Thus $V_{cb}$ and $U_{\mu 3}$
are really not directly related to each other by $SU(5)$; rather
each is related to some {\it right-handed} mixing angle. Specifically,
$SU(5)$ relates $\theta_{sb}^{left}$ to $\theta_{\mu \tau}^{right}$, and
in the toy example both are of order $\epsilon$, as Eqs. (7) and (8) show.
And $SU(5)$ relates $\theta_{\mu \tau}^{left}$ to $\theta_{sb}^{right}$,
and in the toy example both are of order $\sigma$. Moreover,
from the form of the matrices in Eqs. (7) and (8) one sees that the ratio
of masses of the fermions of the second and third families are
given by $m_2/m_3 \sim \sigma \epsilon$. Thus, given the experimental
values of the quark and lepton mass ratios, the values of $\epsilon$
and $\sigma$ are inversely related, so that the smallness of $V_{cb}$ and 
the largeness of $U_{\mu 3}$ are two sides of the same coin.
The expectation from assuming symmetric forms, as in the Fritzschian
textures, for example, would be that the mixing angles 
$\theta_{23}^{Fritzschian}$ are of order $\sqrt{m_2/m_3}$, which
as we just have seen is to say that they are of order
$\sqrt{\sigma \epsilon}$. However, the form of Eqs. (7) and (8) tells
us that actually $V_{cb} \sim \epsilon$ and $U_{\mu 3} \sim \sigma$. 
Thus, as observed, $V_{cb}/\theta_{23}^{Fritzschian} \sim 
\theta_{23}^{Fritzschian}/U_{\mu 3} \sim \sqrt{\epsilon/\sigma} < 1$.

A striking fact about the mechanism just described is that the
large mixing seen in atmospheric neutrino data is coming from a large
off-diagonal entry in $L$, the mass matrix of the {\it charged} leptons.
One is accustomed to speak of ``neutrino mixing angles", but this is
obviously a misnomer, since the leptonic angles are really the
mismatch between the charged and neutral leptons, just as the KM angles
are the mismatch between the up and down quarks.
The mechanism I have just described has three ingredients: (1) There is
large mixing in $L$; (2) $L$ is highly
asymmetric in its 23 block; and (3) $L$ is related by
$SU(5)$ to the transpose of $D$.

Although I have described it in a toy example based on $SU(5)$,
this mechanism first emerged independently in several models
that used various groups.$^{3,4,5,6,7}$ The paper of Sato and Yanagida$^6$
used the group $E_7$ broken down to an $SU(5) \times U(1)$ subgroup.
The paper of Irges, Lavignac, and Ramond used $SU(3) \times SU(2)
\times U(1)^3$, where anomaly cancellation led to $SU(5)$-like
conditions on the mass matrices. Our papers$^{3,4,5}$ were based on
$SO(10)$, which, of course, has $SU(5)$ as a subgroup. Since these first
papers, many papers$^8$ have appeared that consider this idea in models
with the unified group $SU(5)$, while an $SO(10)$ model quite similar in
some respects to the one I am about to discuss has been proposed 
by Babu, Pati and Wilczek.$^9$ In a short talk I cannot go into the details 
of our model. The point I wish to emphasize is that this model was first
constructed without any thought to the pattern of neutrino masses and
mixings that would arise. The goal in our first paper$^3$
was rather to construct a realistic $SO(10)$
model with as simple a Higgs structure as possible. The ``minimal
Higgs structure" in $SO(10)$ involves the use of only the following
non-singlet Higgs representations to break $SO(10)$ down to the
standard model group $G_{SM}$: 
${\bf 45}_H + {\bf 16}_H + \overline{{\bf 16}}_H$.
That the breaking down to $G_{SM}$ can be done with these fields was shown 
in Ref. 10. Starting with this minimal Higgs structure for $SO(10)$
breaking, we looked for the simplest effective Yukawa operators
that can reproduce the pattern of quark and lepton masses that is seen.
By ``simple" operators we mean operators of low dimension,
which can arise from simple tree-level diagrams where small multiplets
are integrated out.

We were led, practically uniquely, to a
set of six Yukawa operators that give the following Dirac mass matrices for
the up quarks, down quarks, neutrinos, and charged leptons at the
GUT scale.$^{3,4}$ (The Majorana mass matrix $M_R$ of the right-handed 
neutrinos comes from different terms.)

\begin{equation}
U= \left( \begin{array}{ccc} \eta & 0 & 0 \\ 0 & 0 & \epsilon/3 \\
0 & - \epsilon/3 & 1 \end{array} \right) m_U,
\end{equation}

\begin{equation}
D = \left( \begin{array}{ccc} 0 & \delta & \delta' \\ \delta & 0
& \sigma + \epsilon/3 \\ \delta' & - \epsilon/3 & 1 \end{array} 
\right) m_D,
\end{equation}

\begin{equation}
N = \left( \begin{array}{ccc} \eta & 0 & 0 \\ 0 & 0 & - \epsilon \\
0 & \epsilon & 1 \end{array} \right) m_U,
\end{equation}

\begin{equation}
L = \left( \begin{array}{ccc} 0 & \delta & \delta' \\ \delta & 0 &
-\epsilon \\ \delta' & \sigma + \epsilon & 1 \end{array} \right) m_D.
\end{equation}

Much about the pattern of entries in these matrices can be understood in 
purely group-theoretic terms. The ``1" entries come from the usual
minimal Yukawa term ${\bf 16}_3 {\bf 16}_3 {\bf 10}_H$. As is well-known,
such a term gives equal contributions to the up quark matrix $U$ and
neutrino Dirac matrix $N$, and it also gives equal contributions to
$D$ and $L$. That is why we can write the matrices in terms of only
two overall scales $m_U$ and $m_D$. The ``$\epsilon$" entries arise
from the lowest dimension Yukawa operator that involves the ${\bf 45}_H$,
namely ${\bf 16}_i {\bf 16}_j {\bf 10}_H {\bf 45}_H$. Because the vacuum
expectation value of the ${\bf 45}_H$ must be proportional to the 
generator $B-L$ in order to do the breaking of $SO(10)$ (in particular
the doublet-triplet splitting), there arises from this term a
relative factor of $- 1/3$ between the quark matrices and the lepton 
matrices, which can be seen in Eqs. (9) --- (12). It can also be
shown that with $\langle {\bf 45}_H \rangle \propto B-L$ the
quartic Yukawa operator involving this VEV gives a flavor-antisymmetric
contribution, as also seen in Eqs. (9) --- (12). The lopsided 
$\sigma$ entries come from a quartic term ${\bf 16}_2 {\bf 16}_3
{\bf 16}'_H {\bf 16}_H$, where the $SO(10)$ indices are contracted
in a certain way. (No effective Yukawa of lower dimension can
be written down that involves the ${\bf 16}_H$.) It is well-know that
such a four-16 operator contributes only to the down quark and charged 
lepton mass matrices. Moreover, since the expectation value of
the spinor ${\bf 16}_H$ breaks $SO(10)$ only down to $SU(5)$, this
operator respects the minimal $SU(5)$-relation $L = D^T$. The
entries $\delta$ and $\delta'$ also arise from four-16 operators, though
ones where the $SO(10)$ indices are contracted differently in a way
that gives flavor-symmetric contributions. However, for the same
reason as in the case of $\sigma$, these only appear in $D$ and $L$.

One cannot in a short talk explain in detail the structure of the model.
Suffice it to say that the mass matrices arise from very simple
Yukawa structures that in turn arise from very simple particle content.
The model is thus not only simple at the level of the mass matrix
``textures", but also at the level of the underlying unified model.

Although it has very few parameters, this model gives a remarkably
good fit to all the quark and lepton masses and mixings.$^4$ There are
altogether {\it nine} predictions. Three of them are well-known
relations that arise in many models because of the group theory
of $SU(5)$ and $SO(10)$: {\bf (1)} $m_b^0 \cong m_{\tau}^0$
(the superscript zero refers throughout to quantities evaluated at the
unification scale); {\bf (2)} $m_s^0 \cong \frac{1}{3} m_{\mu}^0$; 
{\bf (3)} $m_d^0 \cong 3 m_e^0$ (these last two relations are the well-known
Georgi-Jarlskog relations). 

The fourth prediction is that $m_u$ is
relatively small. {\bf (4)} $m_u/m_t \ll m_d/m_b, m_e/m_{\tau}$.
The point is that the Yukawa terms (involving the
parameters $\delta$ and $\delta'$) that generate masses for the other 
fermions of the first family ($d$ and $e$) leave the $u$ quark massless. 
(We have seen the group-theoretical reason for this.) This accords 
well with the fact that $m_d^0/m_b^0 \cong 10^{-3}$, and $m_e^0/m_{\tau}^0
\cong 0.3 \times 10^{-3}$, whereas (assuming $m_u \approx 4$ MeV)
the comparable ratio $m_u^0/m_t^0$ is only about $0.6 \times 10^{-5}$.
Thus a small Yukawa term ($\eta$) must be introduced in this model to give 
a non-vanishing mass for the $u$ quark. This $\eta$ term has interesting
consequences for neutrino masses, as we shall see.

A remarkable postdiction of the model is the charm quark mass$^4$:
{\bf (5)} $m_c(m_c) = (1.1 \pm 0.1)$ GeV. This is remarkable for two
reasons. First, the charm quark mass is generally a severe problem for
$SO(10)$, since the simplest $SO(10)$ schemes predict that $m_c^0/m_t^0
= m_s^0/m_b^0$, which is an order of magnitude too large. Second,
this model not only gives a postdiction of $m_c$ that is of the right
order of magnitude, but even predicts it correctly to within about 15\% 
accuracy, which is quite acceptable given the uncertainties. 

Another remarkable success of the model is a prediction of $V_{ub}$:
{\bf (6)} $V_{ub} = 0.0052 e^{i \theta} - 0.0028$. The phase $e^{i \theta}$
is the single non-trivial physical phase angle contained in the
parameters appearing in Eqs. (9) --- (12). This phase is not fixed by
other measured masses or mixings. So the prediction of the model is
that $V_{ub}$ lies on a certain circle in the complex plane. As it
happens, this circle slices precisely through the middle of the presently
allowed region in the $(Re(V_{ub}), Im(V_{ub}))$ plane. 

Finally, there are predictions for the three neutrino mixing angles.
{\bf (7)} The most significant prediction is that the $\nu_{\mu}-\nu_{\tau}$
mixing is large. This stems from the large entry $\sigma$ in $L$.
A fit to the known quark and lepton masses and mixings gives $\sigma
\cong 1.8$, $\epsilon \cong 0.14$, $\delta \cong 0.008$, $| \delta' |
\cong 0.008$, and $\eta \cong 0.6 \times 10^{-5}$. 
Thus the angle $\theta^{left}_{\mu \tau}$ appearing
in $U_{\mu 3} = \sin (\theta^{left}_{\mu \tau} - 
\theta^{left}_{\nu_{\mu} \nu_{\tau}})$ is given by
$\theta^{left}_{\mu \tau} \cong \tan^{-1} \sigma \cong \pi/3$.
The angle $\theta^{left}_{\nu_{\mu} \nu_{\tau}}$ is not exactly
predicted by the model, since it depends on the unknown Majorana
mass matrix $M_R$ of the right-handed neutrinos. But we can say
from the form of Eq. (11) that it is of order $\epsilon$ and thus
small. 

{\bf (8)} The prediction for the mixing of the electron neutrino is quite
interesting because, depending on what one assumes for $M_R$, it
comes out quite naturally to give {\it either} the small-angle MSW 
solution to the solar neutrino problem {\it or} the vacuum ``just-so"
solution. If one supposes that the matrix $M_R$ does not have large
mixing between the first family and the other families, i.e.
that the 12, 21, 13, and 31 elements of $M_R$ are negligible, then
the mass matrix of the light neutrinos, which has the usual ``see-saw"
form $M_{\nu} = - N^T M_R^{-1} N$, also gives very little mixing between
the first family and the others. (See Eq. (11).) This is also the
case no matter what the form of $M_R$ if the parameter $\eta = 0$
(meaning that $m_u = 0$). In these cases, the mixing of the electron
neutrino comes entirely, or almost entirely, from diagonalizing $L$.
In that case one gets a sharp prediction that $\sin^2 2 \theta_{e \mu}
\cong 16 \times 10^{-3} \cos^2 \theta_{\mu \tau}$. For maximal
mixing of $\nu_\mu - \nu_{\tau}$, as needed to fit the atmospheric
neutrino data, this gives $\sin^2 2 \theta_{e \mu} \cong 8 \times 10^{-3}$,
which is in the allowed range for small-angle MSW. 

On the other hand, if one assumes that $M_R$ has large 12, 21 and/or
13, 31 elements, something quite remarkable happens.
To illustrate, consider the form

\begin{equation} 
M_R = \left( \begin{array}{ccc} 0 & D \epsilon^3 & 0 \\
D \epsilon^3 & B \epsilon^2 & 0 \\ 0 & 0 & A \end{array} \right) m_R,
\end{equation}

\noindent
where we parameterize using powers of $\epsilon$ merely for convenience.
The light neutrino mass matrix comes out to be

\begin{equation}
M_{\nu} = - N^T M_R^{-1} N = \left( \begin{array}{ccc}
\frac{\eta^2}{\epsilon^4} \frac{AB}{D^2} & 0 & - \frac{\eta}{\epsilon^2}
\frac{A}{D} \\ 0 & \epsilon^2 & \epsilon \\ - \frac{\eta}{\epsilon^2}
\frac{A}{D} & \epsilon & 1 \end{array} \right) \frac{m_U^2}{A m_R}.
\end{equation}

\noindent
One sees that the 2-3 block has vanishing determinant, so that a 
rotation in the 2-3 plane by an angle $\theta^{\nu}_{23} \cong 
\epsilon$ brings $M_{\nu}$ to the form

\begin{equation}
M'_{\nu} \cong \left( \begin{array}{ccc}
\frac{\eta^2}{\epsilon^4} \frac{AB}{D^2} & \frac{\eta}{\epsilon}
\frac{A}{D} & - \frac{\eta}{\epsilon^2} \frac{A}{D} \\
\frac{\eta}{\epsilon} \frac{A}{D} & 0 & 0 \\
- \frac{\eta}{\epsilon^2} \frac{A}{D} & 0 & 1 \end{array} \right)
\frac{m_U^2}{A m_R}.
\end{equation}

\noindent
The important thing to notice is that the 1-2 block has a pseudo-Dirac
form. What happens, as a result of this, is that there is almost
exactly maximal mixing between the electron neutrino and the muon
neutrino. In fact, one has ``bimaximal" mixing. Interestingly, the
large value of $U_{\mu 3}$ arises, as we have said, from the {\it charged}
lepton mass matrix, whereas the large value of $U_{e 2}$ is coming, as we
have just seen, from the {\it neutrino} mass matrix. When the parameter
values are looked at more closely, it is found that the vacuum solution
is easily obtained, but the large-angle MSW solution requires 
some fine-tuning of parameters.

{\bf (9)} Finally, there is a prediction for the $\nu_e - \nu_{\tau}$
mixing angle. It is easy to show that in {\it both} cases, the
small-angle MSW case and the bimaximal vacuum oscillation case, 
the mixing $U_{e3}$ comes out the same: $U_{e3} \cong 
0.07 \sin \theta_{\mu \tau}$. 

In conclusion, we have shown that a very simple explanation exists
based on the group theory of $SU(5)$ that accounts for the fact that
the mixing of the left-handed fermions of the second and third
families is small for quarks ($V_{cb} \cong 0.04$) and large for leptons
($\sin^2 2 \theta_{\mu \tau} \cong 1$). We also showed how this
explanation arose from a particular $SO(10)$ model of fermion masses.
This model was seen to be very simple, both at the level of the
underlying $SO(10)$ structures, and at the level of the resulting
mass matrices. Because these matrices involve few parameters,
no fewer than nine predictions result. Several of these will provide
very non-trivial tests of the model, in particular the predictions for
$V_{ub}$ and the mixings $U_{e2}$ and $U_{e3}$. For further details
the reader can consult the series of papers in Refs 4 and 5.

\section*{References}

\noindent
{\small
1. Weinberg, S. {\it Trans. N.Y. Acad. Sci.} {\bf 38}, 185 (1977);
Wilczek, F. and Zee, A. {\it Phys. Lett.} {\bf B70}, 418 (1977);
Fritzsch, H. {\it Phys. Lett.} {\bf B70}, 436 (1977).

\noindent
2. Fritzsch, H. {\it Phys. Lett.} {\bf 73B}, 317 (1978).

\noindent
3. Albright, C.H.  and Barr, S.M. {\it Phys. Rev.} {\bf D58}, 013002
(1998). (9712488)

\noindent
4. Albright, C.H., Babu, K.S. and Barr, S.M. {\it Phys. Rev. Lett.}
{\bf 81}, 1167 (1998) (hep-ph/9802314). Albright, C.H. and Barr, S.M. 
{\it Phys. Lett.} {\bf B452}, 287 (1999).

\noindent
5. Albright, C.H. and Barr, S.M. {\it Phys. Lett.} {\bf B461}, 218 (1999). 

\noindent
6. Sato, J. and Yanagida, T. {\it Phys. Lett.} {\bf B430}, 127 (1998).
(hep-ph/9710516)

\noindent
7. Irges, N., Lavignac, S. and Ramond, P. {\it Phys. Rev.} {\bf D58},
035003 (1998). (hep-ph/9802334)

\noindent
8. Hagiwara, K. and Okamura, N. {\it Nucl. Phys.} {\bf B548}, 60 (1999)
(hep-ph/9811495).
Altarelli, G. and Feruglio, F. {\it Phys. Lett.} {\bf B451}, 388 (1999)
(hep-ph/9812475). Berezhiani, Z. and Rossi, A. hep-ph/9907397.

\noindent
9. Babu, K.S., Pati, J. and Wilczek, F. hep-ph/9812538.

\noindent
10. Barr, S.M.  and Raby, S. {\it Phys. Rev. Lett.} {\bf 79}, 4748 (1997).}

\end{document}